\documentclass[conference]{IEEEtran}
\IEEEoverridecommandlockouts
\usepackage{cite}
\usepackage{amsmath,amssymb,amsfonts}
\usepackage{algorithmic}
\usepackage{graphicx}
\usepackage{textcomp}
\usepackage{xcolor}
\usepackage{cprotect}
\usepackage[numbers]{natbib}
\usepackage{xurl}
\usepackage{tabularx, multirow, booktabs}

\DeclareFontShape{OT1}{cmtt}{bx}{n}{<5><6><7><8><9><10><10.95><12><14.4><17.28><20.74><24.88>cmttb10}{}

\begin{document}

\title{A Robust Learning Rule for Soft-Bounded Memristive Synapses Competitive with Supervised Learning in Standard Spiking Neural Networks}

\author{\IEEEauthorblockN{Thomas F. Tiotto\IEEEauthorrefmark{1}\IEEEauthorrefmark{4},
Jelmer P. Borst\IEEEauthorrefmark{2} and Niels A. Taatgen\IEEEauthorrefmark{3}}
\IEEEauthorblockA{\textit{Groningen Cognitive Systems and Materials Center} \\
\textit{University of Groningen}\\
Groningen, The Netherlands \\
Email: \IEEEauthorrefmark{1}t.f.tiotto@rug.nl,
\IEEEauthorrefmark{2}j.p.borst@rug.nl,
\IEEEauthorrefmark{3}n.a.taatgen@rug.nl\\
\IEEEauthorrefmark{4}Corresponding author}}

\maketitle


\begin{abstract}
Memristive devices are a class of circuit elements that shows great promise as future building block for brain-inspired computing.
One influential view in theoretical neuroscience sees the brain as a function-computing device: given input signals, the brain applies a function in order to generate new internal states and motor outputs.
Therefore, being able to approximate functions is a fundamental axiom to build upon for future brain research and to derive more efficient computational machines.
In this work we apply a novel supervised learning algorithm - based on controlling niobium-doped strontium titanate memristive synapses - to learning non-trivial multidimensional functions.
By implementing our method into the spiking neural network simulator Nengo, we show that we are able to at least match the performance obtained when using ideal, linear synapses and - in doing so - that this kind of memristive device can be harnessed as computational substrate to move towards more efficient, brain-inspired computing.
\end{abstract}


\section{Introduction}
Artificial neural networks (ANNs) are universal function approximators (\cite{hornik1989multilayer}), meaning that - given enough resources (neurons) - an ANN can potentially approximate any and every function to an arbitrary degree of precision.
Philosophically, everything in Nature can be described as a function, including the tasks the brain carries out; for example - alongside all others - those relating to learning, memory, planning, emotions, and self-awareness.
Theoretical Neuroscience has also mostly adopted this view in regards to the brain (\cite{zador2000basic}), defining and studying it as a device which applies functions to its inputs in order to generate outcomes in the form of new internal states and motor outputs.

There nowadays exist modelling tools that explicitly represent brain functionality as mathematical functions with the most well-known and developed probably being the spiking neural network simulator \textit{Nengo} (\cite{bekolay2014nengo}).
Nengo represents information using the principles of the \textit{Neural Engineering Framework} (NEF) (\cite{eliasmith2003neural}), which prescribes how to encode and decode signals into ensembles of spiking neurons and, based on the function being computed, sets out an analytical method to determine the weights of optimal connection matrices between these neuronal pools.
Nengo also implements various \textit{learning rules}, with the MacNeil \& Eliasmith general error-based learning rule \textit{Prescribed Error Sensitivity} (PES) (\cite{macneil2011fine}) is commonly adopted for supervised learning in classic neural network approaches.
Another advantage of Nengo is that is directly compiles to existing digital and mixed-signal neuromorphic hardware, including Intel Loihi, SpiNNaker, FPGAs, and Braindrop.

\textit{Memristors} are a novel class of device that has attracted great research interest since its inception (\cite{chua1971memristor}) and realisation (\cite{strukov2008missing}) due to its capacity to maintain a resistance state in absence of external stimuli.
The fact that the physical state of this device can be directly changed by applying voltage pulses to the terminals allows its programming with a fraction of the power needed for silicon transistors (\cite{jeong2016memristors}).
Memristors have been applied to model neurons, synapses, non-volatile random access memory, and many other more specific applications (\cite{khalid2019review}). 

One of the main reasons to consider a brain-inspired \textit{neuromorphic} approach that uses memristors for computation is because the efficiency improvement that can be gained is prospected to go beyond anything achievable with traditional algorithms running on silicon.
Our present-day machines are struggling to supply us with a sufficient balance between performance and energy consumption (\cite{waldrop2016chips}).
This imbalance is significant when considering the environmental impact that our data centres have when confronted with ever more massive amounts of big data, but is also crucial when the available energy envelope is significantly limited, for example when computing at the extreme edge.
One reason for this relatively poor efficiency can be traced to the fact that contemporary computers are based on the \textit{Von Neumann architecture}, which in itself entails inefficiency, given that memory and computation elements are physically separate within the machine (\cite{efnusheva2017survey}).
Another reason - which follows from the first - is that algorithms written for this computational architecture seem to be inherently less capable of dealing with the kind of uncertainty and massive quantities of data that result from interactions with the real world (\cite{ackley2013beyond,ganguly2019towards}).
In fact, one of the greatest strengths of ANNs is that they do not need to be programmed in the traditional sense, but learn the program/function directly from data.

When looking to advance beyond the Von Neumann architecture and Moore's law, it comes quite natural to look to the brain for inspiration as it is maybe the most remarkable computational device we are aware of.
Our brain consumes only an estimated 20 W of power (\cite{attwell2001energy}), compared to the more than 100 MW needed by a single data centre.
To put things into perspective: the difference in power consumption is the same as if a golf ball were scaled to the size of a medium-sized moon.
Additionally, within this relatively tiny energy envelope the brain is able to solve tasks that our best supercomputers still can't approach (\cite{siegelmann2003neural}).
If we want to start to close the gap between our computers and the brain, we need both new materials and novel computing paradigms: memristors can give us the former, while we can start working towards the latter by adopting a function-based approach as the brain's and ANNs'.

We have already published a paper presenting a novel supervised learning algorithm controlling the resistance of \textit{metal/niobium-doped strontium titanate} (Ni/Nb-doped SrTiO$_3$) memristive synapses, which we named \textit{memristor PES} (mPES), and - as a proof of concept - showed that we could utilise it to learn two very simple functions ($y=x$ and $y=x^2$) (\cite{tiotto2021learning}).
In this work we extend our methodology by applying mPES to learning non-trivial, non-linear functions of varying dimensionality and also compare the results with the optimal analytic implementation of the functions, as computed by Nengo using the NEF principles.
The functions we learn in this work are a lot more interesting than the ones in our original mPES paper (\cite{tiotto2021learning}), as shall be made clear in Section \ref{sec:methods}.  
By implementing our method into Nengo, we show that we are able to at least match the performance of PES; we do so by reproducing and extending the benchmarks by Bekolay (\cite{bekolay2010learning}).

Artificial neural networks use perfectly linear synapses without considering that this might not be the optimal choice, especially in a biologically-plausible \textit{continuous learning} context.
In this context this kind of idealised synapse might, as first noted by \cite{Fusi_Abbott_2007}, actually lead to decreased learning performance and memory capacity in the network.
As shall become clear in the main text, our memristors' resistance dynamics follow a non-linear - power-law - behaviour whose evolution slows as the resistance bounds are approached and this matches the kind of device used by \cite{Brivio_Conti_Nair_Frascaroli_Covi_Ricciardi_Indiveri_Spiga_2018}.
Thus, our current study aims at exploring and confirming the important role that non-linear, soft-bounded synaptic dynamics can play in learning and does so by using a model of a real memristive device in a spiking neural network.
One of the main values of our study is to show that this kind of memristive device can be harnessed as computational substrate to move towards more efficient, brain-inspired computing.

The reader can now see how being able to approximate functions sets the grounds for functional brain simulations and research, which will most certainly be necessary to advance beyond our present-day computational capabilities.
Integrating analogue devices such as memristors into brain-inspired models brings its own set of additional challenges but, given their remarkable energy efficiency, this is quite certainly a valuable undertaking.

\section{Methods}
\label{sec:methods}
In order to compare the performance of our mPES learning rule, to that of standard PES, we reproduce and extend a methodology from existing literature (\cite{bekolay2010learning}).

The PES learning rule (\cite{macneil2011fine}) is incorporated into Nengo and accomplishes online error minimization by solving

\begin{equation}
    \Delta \omega_{i j}=\kappa \alpha_{j} e_{j} \boldsymbol{E} a_{i}
    \label{eq:pes}
\end{equation}

where $\omega_{ij}$ is the weight of the connection between pre-synaptic neuron $i$ and post-synaptic neuron $j$, $\kappa$ is the learning rate, $\alpha_j$ and $e_j$ are NEF-specific parameters for neuron $j$, $\boldsymbol{E}$ is the global $d$-dimensional error to minimize, and $a_i$ the activity of neuron $i$.

\begin{figure}[b]
\centerline{\includegraphics{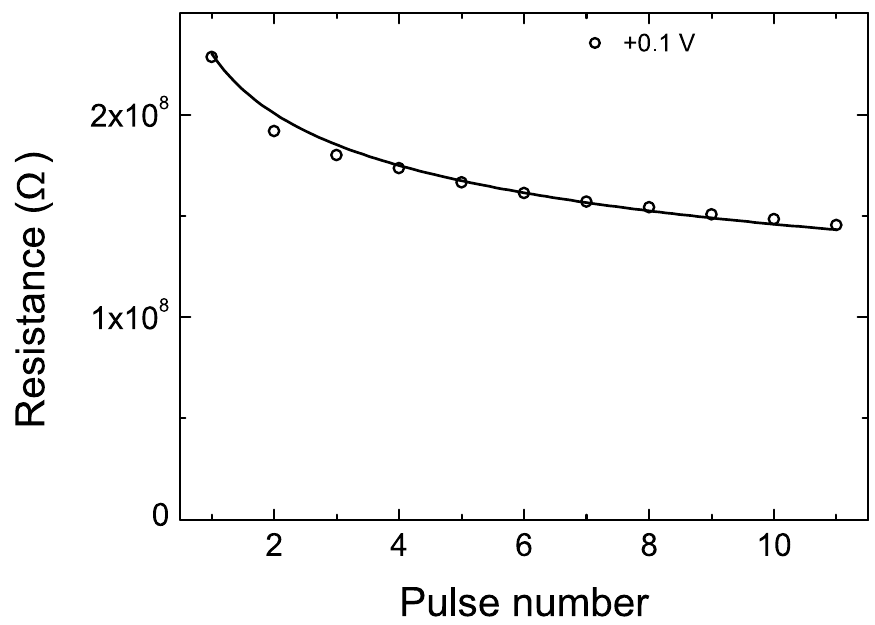}}
\caption{Real memristor data showing the resistance $R(n)$ as a function of the first ten +0.1 V pulses $n$ supplied to the device.
Adapted from \cite{tiotto2021learning}.}
\label{fig:resistance}
\end{figure}

Memristive devices change their internal state and resistance in response to voltages above a threshold.
Here, we utilised simulated Ni/Nb-doped strontium titanate (Ni/Nb-doped SrTiO$_3$) devices, where resistive switching results from changes occurring at the Ni/Nb:STO interface (\cite{goossens2018electric}).
To derive a model of the device, we subjected it to a series of electrical measurements and found that the power-law in Eq. \eqref{eq:powerlaw} could explain the resistance change as a function of the number $n$ of +0.1 V pulses applied to the device terminals

\begin{equation}
\begin{aligned}
R(n) &= R_{0}+R_{1} n^{c} \\
&=200+2.3 \times 10^{8} n^{-0.146}
\end{aligned}
\label{eq:powerlaw}
\end{equation}

with $R(n)$ the device's resistance in Ohm $\Omega$, $R_{0}$ and $R_{0}+R_{1}$ the minimum and maximum resistance, $n$ the number of voltage pulses supplied, and $c$ a constant resulting from fitting.
Fig. \ref{fig:resistance} shows measurements from the real device and highlights how each successive +0.1 V voltage pulses brings the device to a lower resistance state and how each one has a decreasing effect.
The devices can thus be characterised as soft-bound memristive synapses, in line with those explored in existing literature (\cite{Brivio_Conti_Nair_Frascaroli_Covi_Ricciardi_Indiveri_Spiga_2018}).

We do not model a decrease in the resistance due to a lack of experimental data, but this does not impact our rule's capacity to learn.
It is important to note that the initial resistance of each memristor is randomly initialised around a resistance of $10^8 \, \Omega$ and that each update resulting from a pulse also presents a measure of stochasticity in order to better simulate the physical uncertainties one has to deal with when using these kind of materials.
In order to simulate device-to-device variation and hardware noise we introduce variation on both the initial resistance state of the devices and on the magnitude of each update following a voltage pulse.
To this end we initialise the memristor resistances to a random high value in the range $[10^8 \, \Omega \pm 15\%]\, $ to reproduce the case (as measured in the real devices) where the resistance is brought to a high value following the application of multiple RESET voltage pulses.
Secondly, we independently add $15\%$ Gaussian noise to the $R_0$, $R_1$, and $c$ parameters in Eq. \ref{eq:powerlaw} at the start of each simulation run.
This has the effect of making every device respond differently in response to the voltage pulses.
We do not draw the parameter values from a new distribution at each timestep because of computational constraints.
The full details on the device, measurements, and modelling can be found in the Materials and Methods section of our original mPES paper (\cite{tiotto2021learning}).

To be able to represent negative network weights - as \textit{resistance} $R$ and its reciprocal \textit{conductance} $G$ are positive physical quantities - we define the weight $\omega_{ij}$ of each synapse in the Nengo model as the scaled difference between the normalised conductances of two memristors $M^+_{ij}$ and $M^-_{ij}$ as

\begin{equation}
\begin{aligned}
\omega_{ij}=\gamma\left[\left(\frac{G^{+}_{ij}-G^{+}_{0,ij}}{G^{+}_{1,ij}-G^{+}_{0,ij}}\right)-\left(\frac{G^{-}_{ij}-G^{-}_{0,ij}}{G^{-}_{1,ij}-G^{-}_{0,ij}}\right)\right]
\end{aligned}
\label{eq:mpes}
\end{equation}

with $\gamma$ a gain factor (in all simulations in this work fixed at $\gamma=10^3$), and $G^{\pm}_{ij}$, $G^{\pm}_{0,ij}$, and $G^{+}_{1,ij}$ the present, minimum, and maximum conductance of $M^{\pm}$, respectively.

mPES is a novel learning rule that operates on the memristors' conductances $G^{\pm}_{ij}$ in Eq. \eqref{eq:mpes} and is - essentially - a discretised version of PES.
Unlike PES, which operates on ideal weights, mPES needs to account for the non-idealities and discrete changes - which decrease in magnitude with every subsequent pulse - that result from applying voltage pulses to the memristors, as can be seen in Fig. \ref{fig:resistance}.
Each memristor $M^\pm_{ij}$ receives a voltage pulse to increase its conductance based on the sign of the $V=e_j \boldsymbol{E} a_i$ term, defined analogously as in Eq. \ref{eq:pes}, which is a matrix where every entry corresponds to a synapse in the Nengo model.
The \textit{local error} $e_j \boldsymbol{E}$ represents how much the post-synaptic neuron $j$ contributes to the global error $\boldsymbol{E}$, and $a_i$ is the activity of the pre-synaptic neuron.
PES also calculates this the local error, but mPES skips the weight update if it is below a certain threshold because a small local error has little to no effect in a real-valued weight update setting but it can trigger a voltage pulse on a memristor and bring it to a new discrete resistance state, thus having an outsized effect.
If the sign of $V_{ij}$ - corresponding to the synapse between neurons $i$ and $j$ - is positive then the memristor $M^{+}_{ij}$ is pulsed.
Each pulse increases the conductance of $M^{+}_{ij}$ and this, as can be derived from Eq. \ref{eq:mpes}, strengthens the network weight $\omega_{ij}$ i.e., facilitates the synapse between $i$ and $j$.
A symmetric operation is applied to $M^{-}_{ij}$ when $V_{ij}<0$ in order to weaken $\omega_{ij}$.
The overall effect of this procedure is that synapses whose neurons had a beneficial participation to the error are facilitated, and those whose neurons had a negative effect are depressed so as to decrease the probability of them playing a role in the future.
The full details can be found in the Materials and Methods section of our original mPES paper (\cite{tiotto2021learning}).

In order to test the learning performance of mPES, we reproduced the methodology described in ``Learning nonlinear functions on vectors: examples and predictions'' (\cite{bekolay2010learning}) which was used to benchmark PES.
This was done by comparing the quality of multi-dimensional functions $f$ where the matrix implementing the transformation was learned with PES to the case where the matrix entries were computed analytically at the start of the simulation by Nengo using the NEF (referred to simply as ``NEF'' from here onwards). 
In the latter - NEF - case, the matrix implementing the function $f$ is not learned and is not plastic, meaning that it does not change during a simulation run.
We extended this methodology by re-implementing it with the latest version of Nengo at the time of writing (\verb|Nengo 3.0|) and compared the outcome of using our learning rule mPES to those when running PES and NEF.

\begin{figure}[b]
\centerline{\includegraphics[width=\columnwidth]{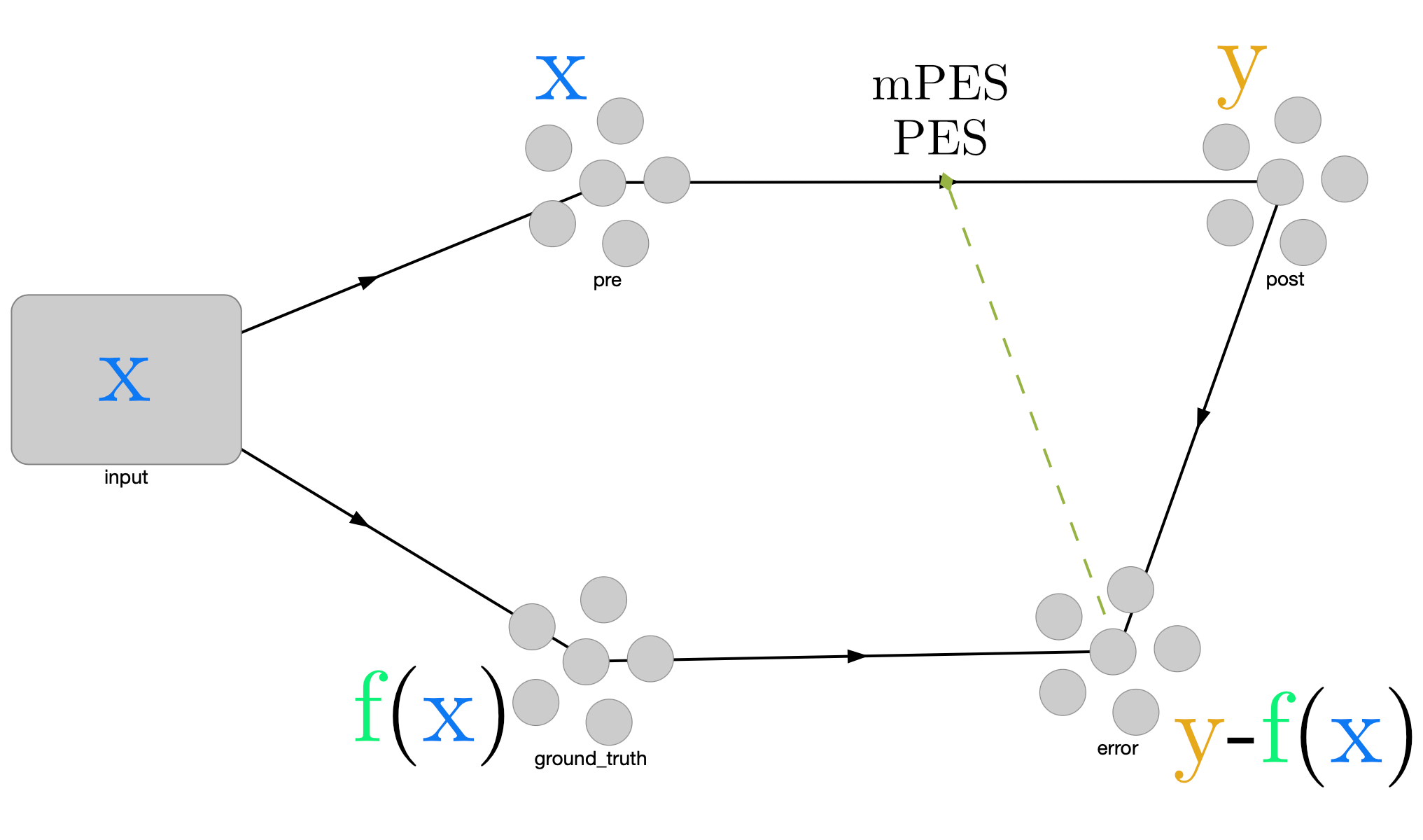}}
\caption{Nengo network topology used to test the learning rule performance. 
The signals $x$, $y$, and $f(x)$ are shown in correspondence to the neuronal ensembles representing them.  
When $f=\otimes$ there is an extra \texttt{conv} ensemble between \texttt{input} and \texttt{ground truth}.
When the NEF is being benchmarked the \texttt{error} ensemble is omitted.}
\label{fig:topology}
\end{figure}

The simple network topology shown in Fig. \ref{fig:topology} is defined specifically to test the learning rule: it consists of a noisy \verb|input| signal $x$ projecting to a \verb|pre|-synaptic neuronal ensemble, a \verb|post|-synaptic ensemble representing $y$ which is connected to \verb|pre| via a plastic connection, a \verb|ground truth| ensemble representing the transformed input $f(x)$, and finally an \verb|error| ensemble which compares the activity in \verb|post| and in \verb|ground truth| as $y-f(x)$.
The input $x$ is an ideal white noise signal in multiple independent dimensions, low-pass filtered by a 5 Hz cutoff.
It is important to note the the function $f$ is computed across the \verb|input|-to-\verb|ground truth| connection analytically by the NEF; learning only happens on the \verb|pre|-to-\verb|post| connection.
The number of neurons in each neuronal ensemble and the simulation run time are all altered depending on the specific function being tested, with the specific values being reported in Table \ref{tab:f}.
The neurons in each ensemble are leaky integrate-and-fire whose parameters are unaltered from the Nengo defaults.

We test learning by running mPES, PES, and NEF ten times each for each of the functions $f$ in Table \ref{tab:f}, which are applied to the \verb|input|-to-\verb|ground truth| connection.
We limit ourselves to ten simulation runs because of computational constraints and also to faithfully reproduce the methodology in (\cite{bekolay2010learning}).
The first function in Table \ref{tab:f} is interesting as an example of the simplest non-linear function.
The second tests if the linear combination of two non-linearities can be learned with a one-dimensional error signal, corresponding to a reinforcement learning setting.
The third function's inclusion is motivated by it being a nearly minimal example of multidimensional-to-multidimensional mapping.
Circular convolution $\otimes$, of the kind of the last two functions in Table \ref{tab:f}, is extremely important in many settings, but especially so in Nengo as it is used to bind vectors together. 
One thing to note is that an extra specialised neuronal ensemble \verb|conv| is required between \verb|input| and \verb|ground truth| in order to compute the circular convolution between $x$'s components.

\begin{table*}[tb]
\scriptsize
\caption{Functions $f$ under test.  
The central columns present the dimensionality $d$ of the signals represented in the main neuronal ensembles together with their number of neurons \#.
The penultimate reports the simulated time for each network run.
The last column shows the error and 95 \% confidence interval measured on the final testing block for mPES, PES, and NEF.}
\begin{center}
\begin{tabularx}{\linewidth}{lccccc}
\toprule
\textbf{Function} $\boldsymbol{f}$ & \texttt{\textbf{Pre (d/\#)}} & \texttt{\textbf{Error (d/\#)}}  & \texttt{\textbf{Post (d/\#)}}  & \textbf{Sim. time} & \textbf{Error (mPES/PES/NEF)} \\
\midrule
$f\left(x_{1}, x_{2}\right)=x_{1} \times x_{2}$ & 2-D / 200 & 1-D / 100 & 1-D / 200 & 50 s & 216 ($\pm 31$) / 227 ($\pm 40$) / 141 ($\pm 20$) \\
$f\left(x_{1}, x_{2}, x_{3}, x_{4}\right)=x_{1} \times x_{2}+x_{3} \times x_{4}$ & 4-D / 400 & 1-D / 100 & 1-D / 400 & 100 s & 394 ($\pm 32$) / 446 ($\pm 60$) / 261 ($\pm 39$) \\
$f\left(x_{1}, x_{2}, x_{3}\right)=\left[x_{1} \times x_{2}, x_{1} \times x_{3}, x_{2} \times x_{3}\right]$ & 3-D / 300 & 3-D / 300 & 3-D / 300 & 100 s & 745 ($\pm 83$) / 773 ($\pm 92$) / 483 ($\pm 60$) \\
$f\left(x_{1}, x_{2}, x_{3}, x_{4}\right)=\left[x_{1}, x_{2}\right] \otimes\left[x_{3}, x_{4}\right]$ & 4-D / 400 & 2-D / 200 & 2-D / 200 & 200 s & 771 ($\pm 70$) / 813 ($\pm 73$) / 710 ($\pm 63$) \\
$f\left(x_{1}, x_{2}, x_{3}, x_{4}, x_{5}, x_{6}\right)=\left[x_{1}, x_{2}, x_{3}\right] \otimes\left[x_{4}, x_{5}, x_{6}\right]$ & 6-D / 600 & 3-D / 300 & 3-D / 300 & 400 s & 1364 ($\pm 64$) / 1317 ($\pm 68$) / 1258 ($\pm 70$) \\
\bottomrule
\end{tabularx}
\label{tab:f}
\end{center}
\end{table*}

The mPES and PES learning rules use the information received from \verb|error| to act on the weights on the plastic connection between \verb|pre| and \verb|post|, with the goal of making the value in \verb|post| resemble that in \verb|ground truth| as much as possible.
This results in the \verb|pre|-to-\verb|post| connection matrix being progressively tuned to represent the transformation $f$ and thus for the value $y$ in \verb|post| to approximate that in \verb|ground truth| i.e., $y \sim f(x)$.
When the NEF is benchmarked - in order to obtain a control ``gold standard'' to gauge mPES and PES's performance - the connection between \verb|pre| and \verb|post| is not plastic but sees its weights pre-calculated in order to optimally represent the function $f$.
In this case, \verb|error| is omitted from the network as $f(x)$ is already optimally being represented on the \verb|pre|-to-\verb|post| connection.

Compared to our original paper (\cite{tiotto2021learning}), we added the \verb|ground truth| and \verb|conv| ensembles.
This made the task more difficult as in the original model $f(x)$ was calculated analytically across the \verb|input|-to-\verb|error| connection while now $f(x)$ is represented by the spiking in \verb|ground truth|.
This extra encoding step leads to a loss of information, making the learning task somewhat harder for mPES to solve.

The learning performance of mPES and PES are compared to the NEF baseline by breaking up a simulation run into an equal number of 2.5 s \textit{learning} and \textit{testing} blocks (for example, the first function in Table \ref{tab:f} sees the networks being run for 50 s, giving a sequence of ten $\|\textit{Learning}\|\textit{Testing}\|$ blocks).
After each simulation run, the values represented by \verb|post| and \verb|ground truth| during each \textit{testing block} are subtracted to give
the absolute total error for that block.
This choice was made because the \verb|error| ensemble is omitted when benchmarking the NEF, but the calculated quantity would be analogous to that represented in \verb|error| during runtime in all cases.

\section{Results and Discussion}
\begin{figure*}[htp]
\centerline{\includegraphics{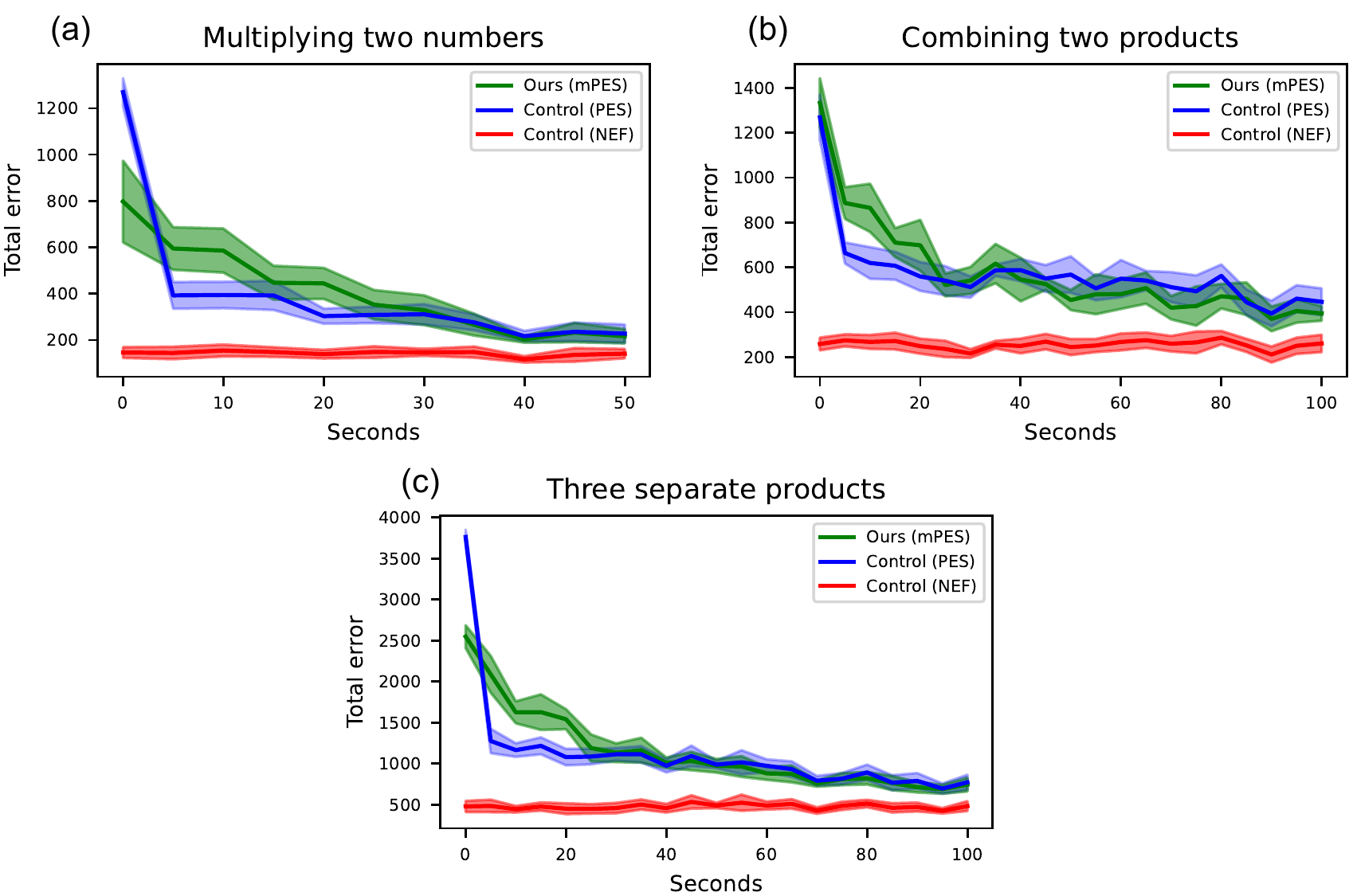}}
\caption{Average accumulated learning error on each 2.5 s testing block for variations on multiplication.
Solid lines represent the accumulated error at each testing block averaged over ten simulation runs and the shaded area corresponds to the 95\% confidence interval for this average value.
\\(a) $f\left(x_{1}, x_{2}\right)=x_{1} \times x_{2}$ \\(b) $f\left(x_{1}, x_{2}, x_{3}, x_{4}\right)=x_{1} \times x_{2}+x_{3} \times x_{4}$ \\(c) $f\left(x_{1}, x_{2}, x_{3}\right)=\left[x_{1} \times x_{2}, x_{1} \times x_{3}, x_{2} \times x_{3}\right]$}
\label{fig:products}
\end{figure*}

\begin{figure*}[htp]
\centerline{\includegraphics{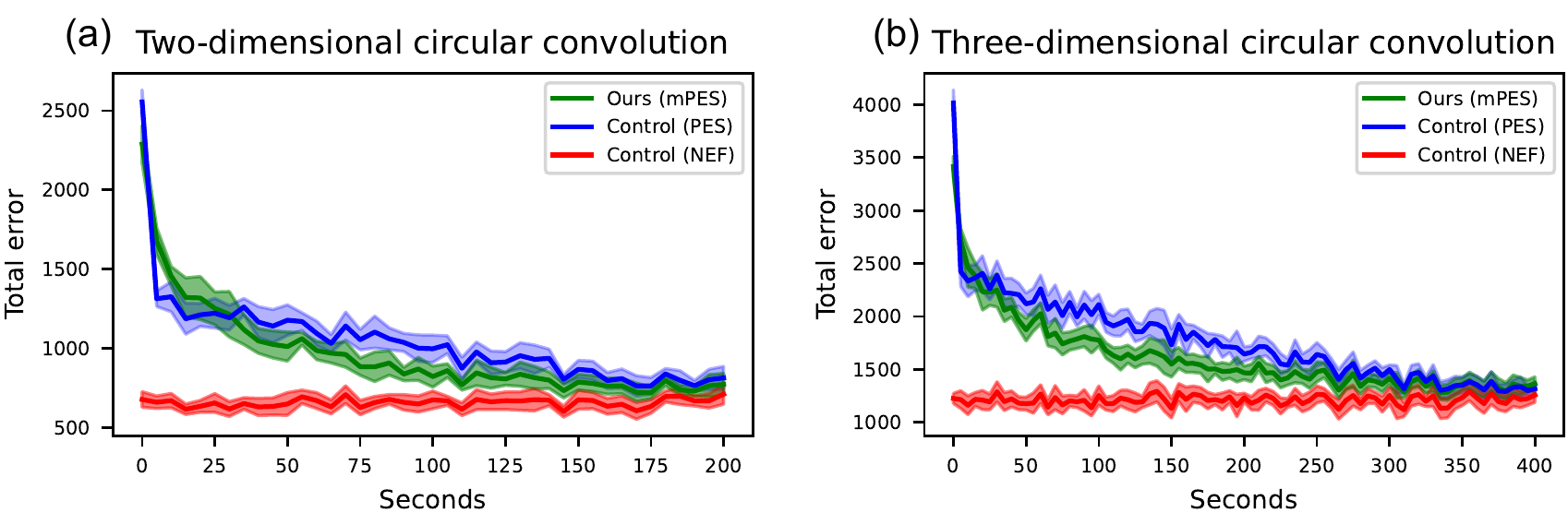}}
\caption{Average accumulated learning error on each 2.5 s testing block for variations on circular convolution.
Solid lines represent the accumulated error at each testing block averaged over ten simulation runs and the shaded area corresponds to the 95\% confidence interval for this average value.
\\(a) $f\left(x_{1}, x_{2}, x_{3}, x_{4}\right)=\left[x_{1}, x_{2}\right] \otimes\left[x_{3}, x_{4}\right]$ \\(b) $f\left(x_{1}, x_{2}, x_{3}, x_{4}, x_{5}, x_{6}\right)=\left[x_{1}, x_{2}, x_{3}\right] \otimes\left[x_{4}, x_{5}, x_{6}\right]$}
\label{fig:convolutions}
\end{figure*}

The results of our simulations comparing mPES to PES and NEF are shown in Fig. \ref{fig:products} and Fig. \ref{fig:convolutions}.
In these plots the solid lines represent the accumulated error at each testing block averaged over ten simulation runs and the shaded area corresponds to the 95\% confidence interval for this average value.
The red colour identifies the NEF control network, blue the control network with PES learning, and green the network using our mPES learning rule.
The $y$-axis shows the average accumulated error during each testing block of the simulation and is in arbitrary units; the $x$-axis shows the total length of the simulation in seconds.
Each data point is spaced 5 s apart, given that the error is reported after a combination of 2.5 s learning and 2.5 s testing blocks ($\|\textit{Learning}\|\textit{Testing}\| \ldots \|\textit{Learning}\|\textit{Testing}\|$).

One might expect the error in the NEF case to present no variation between trials, given that the NEF analytically derives a matrix to optimally represent the function computed across a connection.
In reality, the actual matrix entries in each simulation may differ based on the neuron parameters, which are randomly initialised each time a Nengo simulation is run.
The fluctuations in NEF measured error within a simulation are due to the random nature of the input data and to the transformation matrix better representing some parts of the input space than others.

When learning the five different functions $f$ in Table \ref{tab:f}, mPES was able to modulate the memristive synapses' resistances which, when translated to network weights, implemented $f$ as well or better than PES.
In Fig. \ref{fig:products}a, \ref{fig:products}b, \ref{fig:products}c, where the networks are learning variations on multiplication, mPES sees the total error decrease slower than PES but the error measured on the final testing block is within the confidence interval of PES in all three cases.
When the task is to learn 2-D and 3-D circular convolution of the input components, as shown in Fig. \ref{fig:convolutions}a, \ref{fig:convolutions}b, mPES actually learns the function faster than PES even though the final error is again comparable.
That the performance of mPES is in line to that of PES in all cases, can also be seen by looking at the average error in the last testing block as reported in the rightmost column of Table \ref{tab:f}.
In the mPES and PES cases - where the connection weight matrix is learned - the error at $t=0$ is due to the \verb|pre|-to-\verb|post| matrix being randomly initialised and thus this data point does not reflect the performance of the learning rules.
As noted earlier, the NEF computes the matrix entries analytically based on the specific - random - neuron parameters of that simulation run; for this reason, the performance of the NEF network also shows some variation between runs.
Across the spectrum, as expected, the performance of the NEF analytically-determined weight matrix is superior to that of those obtained via online learning. 

Thus, our mPES learning rule is able to match the learning performance of standard PES, which is remarkable given the restriction imposed by having to operate on non-ideal, stochastic items - as are the simulated memristors - instead of real-valued, linear, continuous network weights.
It is also notable that mPES reaches this level of performance without having any information about the magnitude of the updates happening on the underlying memristors; given that the memristive devices' resistance dynamics follow a power-law subsequent voltage pulses have a monotonically decreasing effect, but mPES is unaware of this.
This leads us to speculate that mPES could also be applied to systems based on different memristive synapses, not just to the Nb-doped SrTiO$_3$ ones utilised in this work, whose behaviour is depicted in Fig. \ref{fig:resistance}.
Additionally the fact that we are still able to at least match the performance of PES (and in the cases reported in Fig. \ref{fig:convolutions} actually do better) supports the finding by \cite{Brivio_Conti_Nair_Frascaroli_Covi_Ricciardi_Indiveri_Spiga_2018} that soft-bounded memristive synapses can enable improved learning performance compared to linear synapses of similar resolution.

In this study we did not directly run ablation experiments to understand the effect of failed devices on the network performance; however, in our original paper (\cite{tiotto2021learning}) we experimented with increasing the variation in the parameters and found that the learning performance of mPES degraded gracefully.
Given that the memristor model follows a power-law (Eq. \ref{eq:powerlaw}) the effect of the exponent $c$ is crucial to the behaviour of the simulated devices: higher variation in $c$ leads an increasing number of devices becoming ``stuck'' in either a high of a low resistance state, effectively rending them unresponsive.

\section{Conclusion}
In this work we have modelled each weight in a spiking neural network by a pair of simulated memristors and shown that a supervised learning rule we call mPES, which operates on these devices' resistances, can match the performance of the standard general error-based learning rule PES (\cite{macneil2011fine}).
The performance of our learning rule has been tested by reproducing and extending a methodology from literature (\cite{bekolay2010learning}), where a Nengo spiking neural network was used as basis for learning the five non-trivial multi-dimensional functions in Table \ref{tab:f}.

Using memristors as basis for neural network weights (and for other ANN elements), especially when paired with a biologically-plausible learning algorithm, is one way of improving the energy efficiency of present-day computing machines.
Memristors are analogue devices and, as such, inherently stochastic; this characteristic, if properly harnessed, could turn out to be important to deal with the randomness present in all data resulting from real-world interactions.
The brain is continually learning from every input it receives and soft-bounded memristive synapses of the kind used in this work can help alleviate the issues of synapse hyper-specialisation and catastrophic forgetting.
\textit{Brittleness} is one of the main downfalls of not just traditional Von Neumann-based algorithms, but of machine learning models as a whole: moving towards more \textit{robust}, \textit{cognitive} systems is taking one step further towards the holy grail of \textit{artificial general intelligence}.

While Nengo directly compiles to certain digital and mixed-signal neuromorphic hardware, we are not currently able to directly implement our model on a crossbar array of SrTiO$_3$ memristors.
Nonetheless, we believe that the main value of this work is in showing that this kind of device can be harnessed as computational substrate to move towards more efficient, brain-inspired computing.

The brain is probably the most extraordinary computational device we know of, but how it carries out its feats of intelligence is still mostly a mystery.
One way of understanding the brain is to view it as a function-computing machine able - for example - to apply a function to the inputs received from our retinae and decide that we are looking at a cat.
Therefore, having a memristor-based neuromorphic system which is able to learn to approximate non-trivial functions - as the ones tested in this work - could prove to be a valuable tool to start to functionally reproduce some of the tasks that the brain seems to carry out so easily and that still elude our best computers and learning models.

\section*{Data Availability Statement}
The experimental datasets generated during the current study are available from the corresponding authors on request. 

\section*{Code Availability}
A reproducible distribution of the code used and the experimental datasets generated during in this study can be found on Code Ocean at  \url{https://codeocean.com/capsule/4587296/tree}.

\bibliographystyle{IEEEtranN}
\bibliography{references}

\end{document}